\documentclass[aps,prl,twocolumn,floats,showpacs]{revtex4}
\usepackage{epsfig}
\usepackage{bm}
\usepackage{latexsym}

\begin{document}

\newcommand{\hide}[1]{}
\newcommand{\tbox}[1]{\mbox{\tiny #1}}
\newcommand{\half}{\mbox{\small $\frac{1}{2}$}}
\newcommand{\sinc}{\mbox{sinc}}
\newcommand{\const}{\mbox{const}}
\newcommand{\trc}{\mbox{trace}}
\newcommand{\intt}{\int\!\!\!\!\int }
\newcommand{\ointt}{\int\!\!\!\!\int\!\!\!\!\!\circ\ }
\newcommand{\eexp}{\mbox{e}^}
\newcommand{\bra}{\left\langle}
\newcommand{\ket}{\right\rangle}
\newcommand{\EPS} {\mbox{\LARGE $\epsilon$}}
\newcommand{\ar}{\mathsf r}
\newcommand{\im}{\mbox{Im}}
\newcommand{\re}{\mbox{Re}}
\newcommand{\bmsf}[1]{\bm{\mathsf{#1}}} 


\title{Superconductor-proximity effect in hybrid structures: Fractality versus Chaos }
\author{Alexander Ossipov$^{1,2}$ and Tsampikos Kottos$^{1,3}$
\\
$^1$Max-Planck-Institut f\"ur Str\"omungsforschung, Bunsenstra\ss e 10,
D-37073 G\"ottingen, Germany\\
$^2$ICTP, Condensed Matter Section, Strada Costiera 11, I-34014 Trieste, Italy\\
$^3$Hellenic Army, KE$\Pi$B Thivas, $5A^{\rm th}$ Artillery Unit, 32200 Thiva, Greece\\
}

\begin{abstract}
We study the proximity effect of a superconductor to a normal system with fractal spectrum.
We find that there is no gap in the excitation spectrum, even in the case where the underlying 
classical dynamics of the normal system is chaotic. An analytical expression for the 
distribution of the smallest excitation eigenvalue $E_1$ of the hybrid structure is obtained.
On small scales it decays algebraically as ${\cal P}(E_1)\sim E_1 ^{-D_0}$, where $D_0$ is 
the fractal dimension of the spectrum of the normal system. Our theoretical predictions 
are verified by numerical calculations performed for various models.
\end{abstract}
\maketitle

During the last years there was considerable interest for the study of the statistical 
properties of hybrid normal-superconductor structures. One of the main outcomes of these 
studies was the prediction that a normal system in the proximity of a superconductor 
acquires characteristics that are typical of the superconducting state. Specifically,
if the underlying classical dynamics of the normal system is chaotic, an energy gap in 
the quasiparticle density of states emerges above the Fermi energy. The mean value 
$\left<E_1\right>$ of the gap was found \cite{MBFB96,VBAB01,JSB03} to be proportional 
to the Thouless energy, 
\begin{equation}
\label{prox}
\left<E_1 \right>\propto E_T = G \delta/4\pi =  M T \delta/4\pi 
\end{equation}
where $\left<\cdots\right>$ indicates an ensemble average, $M$ is the number of transverse 
modes of the point contact between the superconductor and the normal system, $T$ is the 
tunnel probability per mode and $\delta\sim 1/L^d$ is the mean level spacing of the $d$-
dimensional normal system of linear size $L$. The product $G=MT$ is the point contact 
conductance in units of $2e^2/h$. As a matter of fact, in a recent publication \cite{VBAB01} 
the statistical fluctuations of the lowest excited state around the mean-field value $\left< 
E_1\right>$ were studied in the framework of the Random Matrix Theory. It was found that 
the gap distribution is a universal function of the rescaled energy  $x=(E_1-\left< E_1
\right>)/\Delta_g$, in a broad range $|x|\ll M^{2/3}$. The width of the gap distribution 
$\Delta_g \sim M^{1/3}\delta$ is parametrically smaller than the gap size $\left<E_1\right>$ 
but bigger than the mean level spacing $\delta$ of the normal system. In contrast, 
normal systems with integrable classical dynamics do not possess any gap near the Fermi 
energy. Instead, their density of states vanishes linearly near the Fermi level. Thus, 
it was naturally proposed, that the appearance or not of a gap in the excitation spectrum 
of a normal system in the proximity with a superconductor can be used in the studies of 
quantum chaos as a measure for distinguishing classically chaotic systems from integrable 
ones. Apart from the two extreme cases discussed above, a fairly good understanding of
the proximity gap was obtained also for the generic case of systems with mixed classical 
phase space \cite{SB99}. In this case it was found that the excitation gap reduces below 
the value of fully chaotic systems (\ref{prox}).

The investigation of the proximity gap has recently been extended to quantum systems in 
the diffusive regime \cite{OSF02} where it was found that a similar type of gap appears. 
The value of $E_1$ is given by Eq.~(\ref{prox}) provided that we substitute the ballistic 
conductance $MT$ with the appropriate expression $DL^{2-d}$ for diffusive systems. Here 
$D$ is the classical diffusion coefficient. 

Despite all this activity, a significant class of systems was left unexplored. Namely 
hybrid structures whose normal part has fractal spectra. The latter exhibit energy 
level statistics that are in strong contrast to the level repulsion predicted by Random 
Matrix Theory (RMT) \cite{P65}. Their level spacing distribution follows inverse power
laws $P(s)\sim s^{-\beta}$ which is a signature of level clustering. The power $\beta$ 
was found to be related with the fractal dimension of the spectrum $D_0$ as $\beta = 
1+ D_0$ \cite{GKP95}. Realizations of this class are quasi-periodic systems with 
metal-insulator transition at some critical value of the on-site potential like the 
Harper model \cite{GKP95,AA80}, Fibonacci chains \cite{GKP95,SBGC84}, or quantum systems 
with chaotic classical limit as the Kicked Harper Model \cite{GKP91,ACS92,ZZSUC86}. Apart 
of their own interest  the analysis of these systems can be illuminating for the 
understanding of the behavior of high dimensional disordered systems
at the metal-insulator transition like the 3$d$ Anderson model \cite{A58}.

Here, for the first time, we present consequences of the fractal nature of the spectrum 
of the normal system in the excitation spectrum of the hybrid structure. We consider 
the normal system  connected to the superconductor via point contacts supporting 
$M$ channels and show that there is no gap in the excitation spectrum even in the case 
where the corresponding classical phase space is chaotic (like in the case of the 
Kicked Harper model). We derive an analytical expression for the distribution of the 
minimum excitation eigenvalue ${\cal P}(E_1)$, and show that its behavior is dictated 
by the fractal dimension $D_0$ of the spectrum of the normal part. Thus {\it the nature 
of the classical dynamics becomes totally irrelevant} for these type of systems. 
Specifically we show that ${\cal P}(E_1)$ generated over different Fermi energies 
behaves as
\begin{equation}
\label{peg}
{\cal P}({\tilde E}_1)=({1\over D_0}-1)({\tilde E}_1^{-D_0} -1)
\end{equation}
where ${\tilde E}_1 = 2 {E_1\over s_{\rm max}}$ and $s_{\rm max}$ is the maximum level
spacing of the normal system within the energy interval that is used in order to generate
 statistics. Eq.~(\ref{peg}) is the main outcome of our investigation. A consequent
result is that the mean ${\tilde E}_1$ is given by
\begin{equation}
\label{meanEg}
\left<{\tilde E}_1\right> = {1-D_0\over 2-D_0}
\end{equation}
One has to note the lack of any dependence on the system size $L$ and the number of channels 
$M$ in contrast to Eq.(\ref{prox}). Our theoretical results (\ref{peg},\ref{meanEg}) are 
confirmed by numerical calculations performed for various systems with fractal spectra.

Let us start our analysis with the Kicked Harper (KH) model. The system is defined by the time 
depended Hamiltonian
\begin{equation}
\label{KH}
H = Q \cos( p) + K \cos(\theta) \sum_m \delta (t-mT)
\end{equation}
where $p$ denotes the angular momentum, $\theta$ the conjugate angle, while the kick 
period is $T$. Contrary to the standard Harper model (corresponding to the limit 
$K\rightarrow 0$) this model for large enough $K\ge 5$ is classically chaotic.

The quantum mechanics of this system is described by a  time evolution operator for one 
period
\begin{equation}
\label{Uop}
U=\exp(-i{Q\over \hbar} \cos(\hbar{\hat p})) \exp(-i{K\over \hbar}\cos(x))
\end{equation}
where ${\hat p} = -id/d\theta$ is the momentum operator and $\hbar$ is an effective
Planck constant, which includes the frequency ratio of the unperturbed system and the
external driving. For $K=Q$ the quasi-energy spectrum is fractal \cite{GKP91,ACS92} and we 
always consider cases where $\hbar/2\pi$ is strongly irrational. Using a recently 
proposed recipe \cite{JSB03} we can write down the corresponding quantum Andreev map
${\cal F}$ and find the quasi-energies of the excitation spectrum of the hybrid 
structure by direct diagonalization of ${\cal F}$. In all cases considered bellow we
have generated more than 3000 values of $E_1$ for statistical processing. 

\begin{figure}
\epsfig{clip,figure=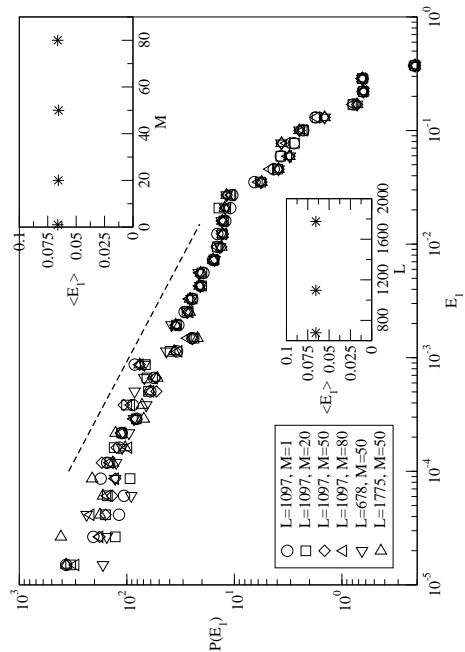,scale=0.35,angle=0}
\caption{
The distribution of the  lowest excited energy, generated over an ensemble of different  Fermi 
energies, for the KH model (\ref{KH}). Various symbols correspond to different system sizes 
and numbers of channels. All data overlap with one another indicating that ${\cal P}(E_1)$
is insensitive to the number of channels and the system size. The dashed line is the theoretical
prediction of Eq.~(\ref{peg}). In the insets we show the numerically evaluated $\left<
E_1\right>$  versus the system size $L$ and the number of channels $M$. 
}
\end{figure}

Figure 1 shows ${\cal P}(E_1)$ for various system sizes $L$ and number of channels $M$.
For small values of $E_1$ the distribution of the minimum excitation eigenvalue displays 
clearly an inverse power law. Moreover, it is independent of the number of channels and 
system size in agreement with Eq.~(\ref{peg}). In the insets of Fig.~1 we also report our 
results for the mean value of $E_1$ for various system sizes and various numbers of channels. 
The inverse power law character of the distribution ${\cal P}(E_1)$ forces us to conclude 
that the probability to find a quasi-energy excitation smaller than $\left<E_1\right>$ is 
high and thus there is no gap  in the excitation spectrum (even in a probabilistic sense). 

The validity of Eqs.~(\ref{peg},\ref{meanEg}) can be verified in more cases in the Fibonacci 
chain model of a one dimensional quasi-crystal where various scaling exponents $D_0$ can be 
obtained. The normal system is described by the tight-binding Hamiltonian:
\begin{equation}
\label {tight-binding}
H=\sum_n |n\rangle V_n\langle n| + \sum_n \left( |n\rangle \langle n+1| +
|n+1\rangle \langle n|\right)
\end{equation}
where $V_n$ is the potential at site $n$. It only takes the two values $+V$ and $-V$ arranged 
in a Fibonacci sequence \cite{SBGC84}. It was shown that the spectrum is a Cantor set with zero
Lebesgue- measure for all $V>0$. The sample is in contact with $M$ semi-infinite one-dimensional 
superconductors which are attached in $M$ randomly chosen sites. The quasi-energy spectrum of 
the hybrid structure is calculated by employing the effective Hamiltonian approach \cite{MBFB96}.
We again find inverse power laws for the distributions ${\cal P}({E}_1)$ . Here the
exponent depends on the potential strength $V$, while Eq.~(\ref{peg}) still relates the resulting
statistics to the fractal dimension $D_0^E$. 

In Fig.~2 we report our results for ${\cal P}(E_1)$ for $V=1.4$ and various system sizes $L$ 
and number of channels $M$. In the insets we also plot the mean value $\left<E_1\right>$ of 
the distribution. Similarly with Fig.~1 we observe that ${\cal P}(E_1)$ and consequently 
$\left< E_1\right>$ are independent of $L$ and $M$. In Fig.~3 we summarize our results for 
various $V$ values. A nice agreement between our numerical data and the theoretical predictions 
(\ref{peg},\ref{meanEg}) is observed \cite{note1}.

\begin{figure}
\epsfig{clip,figure=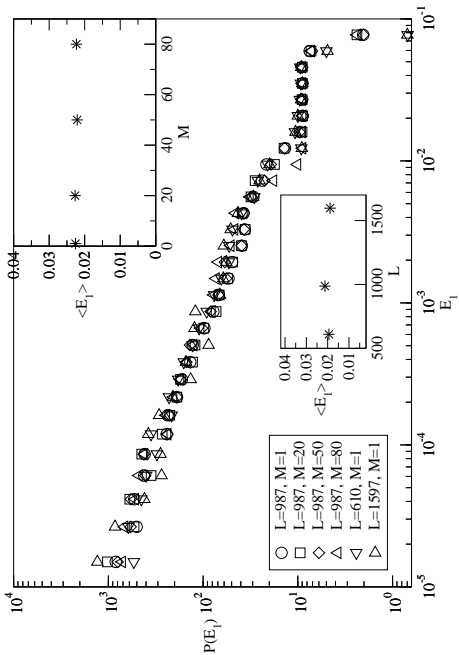,scale=0.35,angle=0}
\caption{
The distribution ${\cal P}(E_1)$ of a hybrid structure consisting of a a Fibonacci sample of size 
$L$ (normal part) attached with $M$ semi-infinite one-dimensional superconductors. The data for
various $L,M$ fall one on top of the other, indicating that ${\cal P}(E_1)$ is independent of
$L,M$. In the insets we report the mean value $\left<E_1\right>$ of the distribution as a function 
of system size $L$  and number of channels $M$.
}
\end{figure}
%

The above results call for a theoretical explanation. Our starting point is the observation 
that the distribution ${\cal P}(E_1)$ rescaled in appropriate way is the same for the normal 
and the hybrid structure. This assumption is verified in Fig.~4 where we plot the distribution 
of the minimum excitation level $\tilde E_1$ for a representative case where the normal part 
is a Fibonacci lattice with $V=1.4$. The overlap between the resulting distributions of the 
hybrid structure and of the corresponding normal system is evident. In order to understand
this phenomenon one should recall that the eigenstates of the normal system are not extended 
like typical chaotic eigenstates, but they have fractal structure. As a result, most of the 
eigenstates have intensities at the boundary with a superconductor which are so small that 
one can consider that they are practically not affected by the proximity of the system to 
the superconductor. Thus these eigenstates and the corresponding eigenenergies are solutions 
of the eigenvalue problem for  the hybrid structure as well. We point here that the perturbative
 assumption used in our argument, was verified numerically in \cite{OWKG01}, where it was observed 
that the effect of the coupling of systems with fractal spectra to external leads is a small 
perturbation for the most of the eigenstates.

\begin{figure}[t]
\epsfig{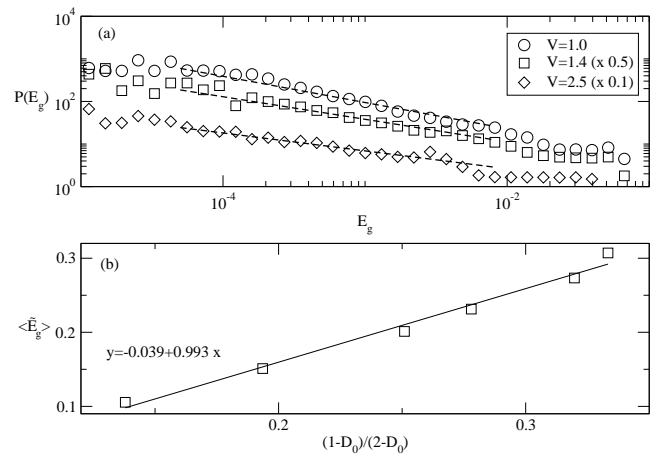}
\caption{
(a) The distribution ${\cal P}(E_1)$ for hybrid structures where the normal sample is a Fibonacci 
lattice with $L=987$ and $M=1$ and various $V$'s. The dashed lines are the theoretical predictions 
(\ref{peg}); (b)  The mean value  $\left<\tilde E_1\right>$  versus the theoretical prediction 
(\ref{meanEg}) for various $V$ values. The straight line $y=-0.039+0.993x$ represents the best 
linear fit.
}
\end{figure}

The distribution ${\cal P}(E_1)$ for the normal system  can be
derived in the following way. We scan the spectrum of the normal system with a set
of Fermi energies which have resolution given by $\epsilon$. Then for fixed $E_1$ we count 
the number of Fermi energies that are in a distance $E_1$ with tolerance $\epsilon$ from
the next larger eigenenergy of the normal system. This is given by the number of level
spacings which are larger than $E_1$ i.e. $\int_{E_1}^{s_{\rm max}} p(s) ds$ where $s_{\rm 
max}$ is the maximum level spacing and $p(s)$ is the level spacing distribution. The
normalized gap density in the limit of $\epsilon\rightarrow 0$ is then given by
\begin{equation}
\label{theo1}
{\cal P}(E_1) = {\int_{E_1}^{s_{\rm max}} p(s) ds \over 
\int_0^{s_{\rm max}} dE_1 \int_{E_1}^{s_{\rm max}} p(s) ds}
\end{equation}
Substituting in the above equation the expression for $p(s)=s^{-(1+D_0)}$ we eventually
get
\begin{equation}
\label{theo2}
{\cal P}({\tilde E}_1)=({1\over D_0}-1)({\tilde E}_1^{-D_0} -1)
\end{equation}
where ${\tilde E}_1 = {E_1\over s_{\rm max}}$ is the rescaled energy gap. Notice that an 
additional factor $2$ should be introduced in the definition of the rescaled energy gap
i.e. ${\tilde E}_1 = 2 {E_1\over s_{\rm max}}$ once we turn to the distribution of the
hybrid structure, due to the fact that the excitation
levels for the hybrid structure are coming in pairs. In Figs.~1 and 3a we plot with dashed 
lines the above theoretical prediction. Using Eq.~(\ref{theo2}) the mean $E_1$ and can be easily 
evaluated leading to  Eq.~(\ref{meanEg}).

\begin{figure}
\epsfig{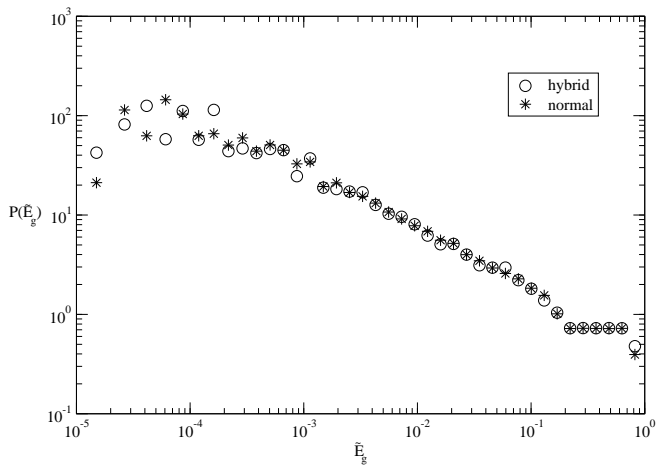}
\caption{
The distribution ${\cal P}({\tilde E}_1)$ for a hybrid structure where the normal part is a Fibonacci 
system with $L=987$ and $M=50$ and $V=1.4$ ($\circ$). Overplotted ($\star$) is the corresponding
distribution ${\cal P}({\tilde E}_1)$  for the same Fibonacci system disconnected from a superconductor. 
}
\end{figure}

In conclusion we have studied the statistical properties of the excitation spectrum of a normal
system with fractal spectra in the proximity to a superconductor. We have found that the underlying
classical dynamics is irrelevant and that the statistical properties of the quasi-spectrum depends
only on the fractal nature of the normal system. Such a system can be realized by a two-dimensional 
electron gas subject to a perpendicular magnetic field and periodic potential in contact with a 
superconductor \cite{2DEG}. It is known that the normal system of this type can be mapped onto
the Harper model \cite{GKP95,AA80,note1} possessing fractal spectrum at the critical point. 

Finally, our results can be of interest for the quantum optics community with respect to studies 
of reflection of light by a dielectric medium in front of a phase-conjugation mirror 
\cite{PBB97}. This problem is the optics analogue of Andreev reflection \cite{note2}. A wave 
incident at frequency $\omega_0 + \Delta \omega$ is retro-reflected at frequency $\omega_0 - 
\Delta \omega$ where $\omega_0$ is the pumped frequency of the phase-conjugation mirror. The 
analogue of $\omega_0$ and of the frequency shifts $\Delta \omega$ are the Fermi energy $E_F$ 
and the excitation energies $E$ respectively. The latter were shown here to be controlled by 
the fractal dimension $D_0$. Keeping in mind the above analogies it is reasonable to conjecture 
that $\Delta \omega$ is controlled as well by $D_0$ in quasi-periodic optical structures like
e.g. in Fibonacci quasicrystals \cite{Negro03}. Thus controlling $D_0$ we can tune $\Delta 
\omega$ which was predicted in \cite{PBB97} to affect drastically the reflected intensity from 
a phase conjugation mirror for a certain parameter range. 

It is our pleasure to thank J. Cserti, R. Fleischmann, C. Lambert and H. Schomerus, for 
useful discussions. T.K.  acknowledges the officers of the $5^{\rm th}$A Artillery Unit 
for their support. This research was supported by a grant from the GIF, the German-Israeli 
Foundation for Scientific Research and Development.



\begin{thebibliography}{99}
\bibitem{note1} We have also checked the validity of Eqs.~(\ref{peg},\ref{meanEg}) 
for the Harper model at criticality described by Eq.~(\ref{tight-binding}) with
$V_n=2 \cos(2\pi\sigma n)$ and $\sigma=({\sqrt 5}+1)/2$.
\end{thebibliography}
\end{document}